\input psfig.sty
\def\ho{{$^1$H}}
\def\het{{$^3$He}}
\def\hef{{$^4$He}}

\def\pn{{\par\noindent}}
\def\thn{{\thinspace}}
\def\scr{\scriptstyle}
\def\ref{\par \noindent \hangindent=3pc \hangafter=1}

\def\Msun{\hbox{$\thn M_{\odot}$}}

\def\Lsun{\hbox{$\thn L_{\odot}$}}
\def\={\thn\thn=\thn\thn}

\def\tgs{{\thn \rlap{\raise 0.5ex\hbox{$\scr  {>}$}}{\lower 0.3ex\hbox{$\scr  {\sim}$}} \thn }}
\def\tls{{\thn \rlap{\raise 0.5ex\hbox{$\scr  {<}$}}{\lower 0.3ex\hbox{$\scr  {\sim}$}} \thn }}
\def\tll{{\raise 0.3ex\hbox{$\scr  {\thn \ll \thn }$}}}
\def\tgg{{\raise 0.3ex\hbox{$\scr  {\thn \gg \thn }$}}}
\def\tle{{\raise 0.3ex\hbox{$\scr  {\thn \le \thn }$}}}
\def\tge{{\raise 0.3ex\hbox{$\scr  {\thn \ge \thn }$}}}
\def\tl{{\raise 0.3ex\hbox{$\scr  {\thn < \thn }$}}}
\def\tg{{\raise 0.3ex\hbox{$\scr  {\thn > \thn }$}}}
\def\ts{{\raise 0.3ex\hbox{$\scr  {\thn \sim \thn }$}}}

\def\do{{\tt "}}
\def\tp{{\raise 0.3ex\hbox{\small +}}}

\def\z{\ \ }
\def\dm{\Delta\mu}

\documentclass[12pt]{article}

\usepackage{scicite}

\usepackage{times}

\newenvironment{sciabstract}{%
\begin{quote} \bf}
{\end{quote}}

\newcounter{lastnote}

\title{Deep Mixing of $^3$He: Reconciling Big Bang and Stellar Nucleosynthesis}

\author
{Peter P. Eggleton$^{1,3\ast}$, David S. P. Dearborn,$^{2,3}$ \break \and John. C. Lattanzio$^4$\\
\\
\normalsize{$^1$Institute of Geophysics and Planetary Physics}\\
\normalsize{$^2$Physics and Applied Technologies Division}\\
\normalsize{$^3$Lawrence Livermore National Lab., 7000 East Ave, Livermore, CA94551, USA}\\
\normalsize{$^4$Centre for Stellar and Planetary Astrophysics, Monash University,
Australia}\\
\\
\normalsize{$^\ast$To whom correspondence should be addressed; E-mail: ppe@igpp.ucllnl.org}
}

\date{}

\begin{document}


\baselineskip24pt


\maketitle


\begin{sciabstract}

Low-mass stars, $\ts 1-2$ solar masses, near the Main Sequence are efficient 
at producing \het, which they mix into the convective envelope on the giant 
branch and should distribute into the Galaxy by way of envelope loss. This 
process is so efficient that it is difficult to reconcile the low observed 
cosmic abundance of \het\ with the predictions of both stellar and Big Bang 
nucleosynthesis. In this 
paper we find, by modeling a red giant with a fully three-dimensional 
hydrodynamic code and a full nucleosynthetic network, that mixing arises in 
the supposedly stable and radiative zone between the hydrogen-burning shell 
and the base of the convective envelope. This mixing is due to Rayleigh-Taylor 
instability within a zone just above the hydrogen-burning shell, where a 
nuclear reaction lowers the mean molecular weight slightly. 
Thus we are able to remove the threat that \het\ production in low-mass stars 
poses to the Big Bang nucleosynthesis of\ \ \het.

\end{sciabstract}


\par The standard evolution of a low-mass star (Fig. 1) takes it from a
short-lived pre-Main-Sequence (PMS) state, in which it contracts and
heats up but has not yet become hot enough to burn its nuclear fuel,
to the long-lived MS state in which slow steady nuclear reactions keep
the star in thermal equilibrium. After several gigayears (but depending
strongly on mass) the nuclear fuel is exhausted at and near the center,
the star becomes cooler, larger and more luminous, and it starts to climb
the Red-Giant Branch (RGB). Its outer layers become turbulent and 
convective, and this Surface Convection Zone (SCZ) penetrates deeply
into the star; but the SCZ is forced to retreat again as the fuel-exhausted
core, surrounded by a thin hot nuclear-burning shell, advances outwards.
During the growing phase the SCZ
dredges up and homogenises material that, at the earlier MS phase, was 
processed by nuclear reactions in the interior. 

\vskip 0.2truein
\centerline{\psfig{figure=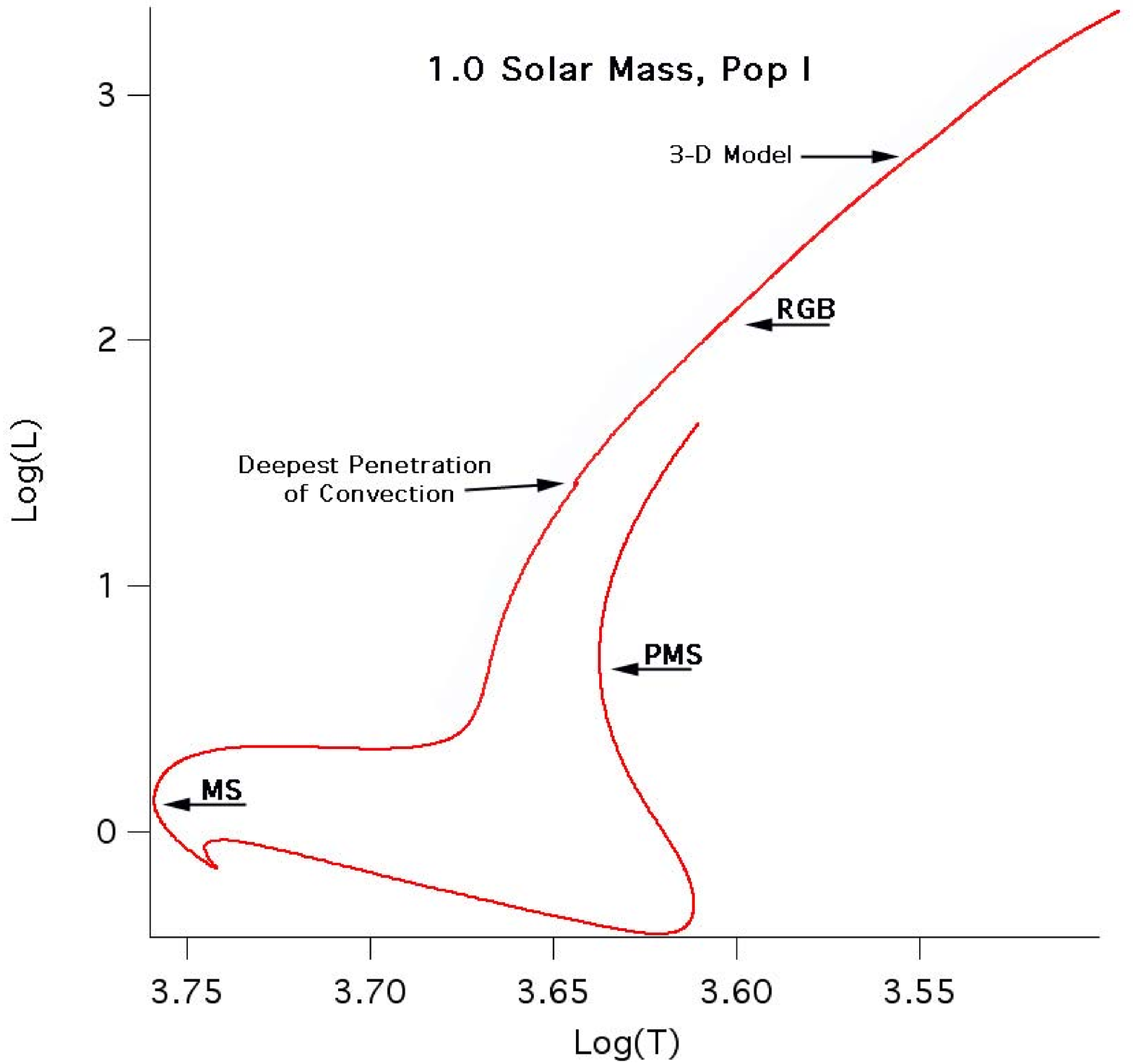,height=3.0in,bbllx=0pt,bblly=0pt,bburx=750pt,bbury=750pt,clip=}}
{\small {\bf Figure 1}. Evolution of a low-mass Pop I star in a 
luminosity-temperature diagram. The model was computed in 
1D, i.e. spherical symmetry was assumed, using the code of ({\it 1, 2})
with updated equation of state, opacity and nuclear reaction rates ({\it 3}). 
Surface temperature is in Kelvins, luminosity in Solar units.}

\par  Along the MS stars burn hydrogen in their cores by a combination 
of the pp chain (in which 4 protons unite to form a \hef\ nucleus)
and the CNO tri-cycle (in which the same process is catalysed by carbon,
nitrogen and oxygen). The former is the more important in low-mass stars,
$\tls 1\Msun$, and the latter in more massive stars. However even in the
more massive stars there is still a shell, somewhat outside the main
energy-producing region, where the the pp chain partially operates, burning
H to \het\ but not beyond.
\par Because the pp chain is less sensitive to temperature than the CNO 
cycle, cores of low-mass stars are free of convection; but convective cores 
develop in higher-mass stars because CNO energy production is too 
temperature sensitive for radiation to stably transport the energy. Above 
$\ts 2\Msun$ this convective core is large enough that the \het\  produced 
is convected into the center of the star and burnt there. But in stars of 
lower mass \het\ accumulates ({\it 4}) in a broad zone outside the main 
energy-producing region (Fig. 2a).  \het\ is enriched above its assumed 
initial value ($2.10^{-4}$ by mass ({\it 5)}, the same as its surface value 
in this plot) in a broad peak extending over nearly half the mass (as well 
as about half the radius) of the star. The maximum \het\ 
abundance in this peak is a factor of $\ts 18$ larger than the initial value.

\vskip 0.1truein
{\psfig{figure=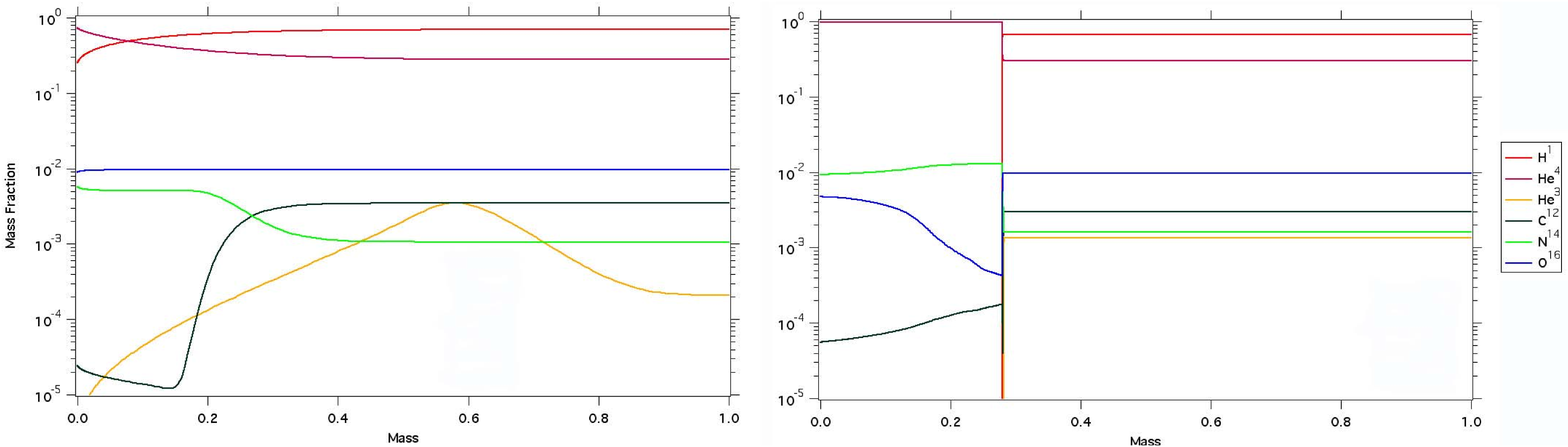,height=2.3in,bbllx=0pt,bblly=0pt,bburx=2050pt,bbury=750pt,clip=}}
\vskip 0.2truein
{\small {\bf Figure 2}.  (a) Profiles of the abundances of certain nuclei in 
a star which has evolved to roughly the end of the MS (Fig. 1; 
$T\ts 5000\thn$K, $L\ts 2\Lsun$). $^1$H is orange, $^4$He is red,
$^{16}$O is blue, $^{12}$C is black, $^{14}$N is green and \het\ is yellow.
\het\ shows a major peak where the abundance reaches $\ts 18$ times 
the initial (surface) abundance. (b) The same star later, when the SCZ 
reaches its maximum inward extent (Fig 1).
The \het\ peak has been homogenized, to a factor of 8 larger than its initial
value. The inert, H-depleted core is about $0.27\Msun$.}
\vskip 0.1truein
 
\par On the lower part of the RGB (Fig. 1), a large SCZ develops, which
mixes and homogenises the outer $\ts 0.7\Msun$ (Fig. 2b). The surface \het\ 
abundance is raised from the initial of $2.10^{-4}$ to $\ts 1.6.10^{-3}$, 
i.e. by a factor of $\ts 8$. 

\par As the star climbs the RGB beyond the point (Fig. 1) where the SCZ
penetrates most deeply, the
SCZ is diminished by (a) nuclear burning below its base, in a zone
that marches outwards, and (b) stellar-wind mass loss from its surface.
The evidence for the latter is that the next long-lived stage after the RGB
is the Horizontal Branch (HB), and HB stars appear to have masses that are
typically $0.5 - 0.6\Msun$, substantially less than the masses of stars
capable of evolving to the RGB in less than a Hubble time ({\it 6, 7}). 
Process (b) leads to enrichment of the interstellar medium (ISM) in \het\ 
({\it 8, 9, 10}).

\par Yet the ISM's abundance of \het, at $\ts 5.10^{-5}$ by mass, is little 
different from that predicted by Big Bang nucleosynthesis. This is a major 
problem ({\it 11, 12}): either the Big Bang value is too high, or the
evolution of low-mass stars is wrong. 
\par In this paper we identify a mechanism by which low-mass stars destroy
(on the RGB) the \het\  that they produced during their MS evolution. Although
we illustrate this with a star like the Sun, regarding both mass and initial
composition, we emphasise that exactly the same applies to low-mass metal-poor
stars (`Population II'), which may have been more important than metal-rich
(`Population I') stars like the Sun throughout the earlier part of Galactic 
history in determining the \het\ abundance of the interstellar medium. The
process is largely independent of mass provided it is fairly low: $1-2\Msun$
for Pop I and $0.8-1.6\Msun$ for Pop II.

\par Once the SCZ has reached its deepest extent, part-way up from the
base of the RGB, it retreats, and can be expected to leave behind a region
of uniform composition with \het\ enhanced (Fig. 2b). This region is stable 
to convection according to the usual criterion that the temperature gradient 
should be sub-adiabatic, and is quite extensive in radius although small in 
mass. The H-burning front moves outwards into the stable region, but preceding 
the H-burning region proper is a narrow region, usually thought unimportant, 
in which the \het\ burns. The reaction that mainly consumes it is 
$^3{\rm He}\ (^3{\rm He}, 2{\rm p})^4{\rm He}\ ,$
which is an unusual reaction in stellar terms because it lowers the mean 
molecular weight: two nuclei become three nuclei, and the mean mass per 
nucleus decreases from 3 to 2. Since the molecular weight ($\mu$) is the 
mean mass per
nucleus, but including also the much larger abundances of \ho\ and \hef\ that
are already there and not taking part in this reaction, this leads to a small 
inversion in the $\mu$-gradient. The inversion is tiny (Fig. 3): it is in 
about the fourth decimal place. But our 3D modeling shows the inversion 
to be hydrodynamically unstable, as we should expect from 
the classic Rayleigh-Taylor instability.

\vskip 0.2truein
\centerline{\psfig{figure=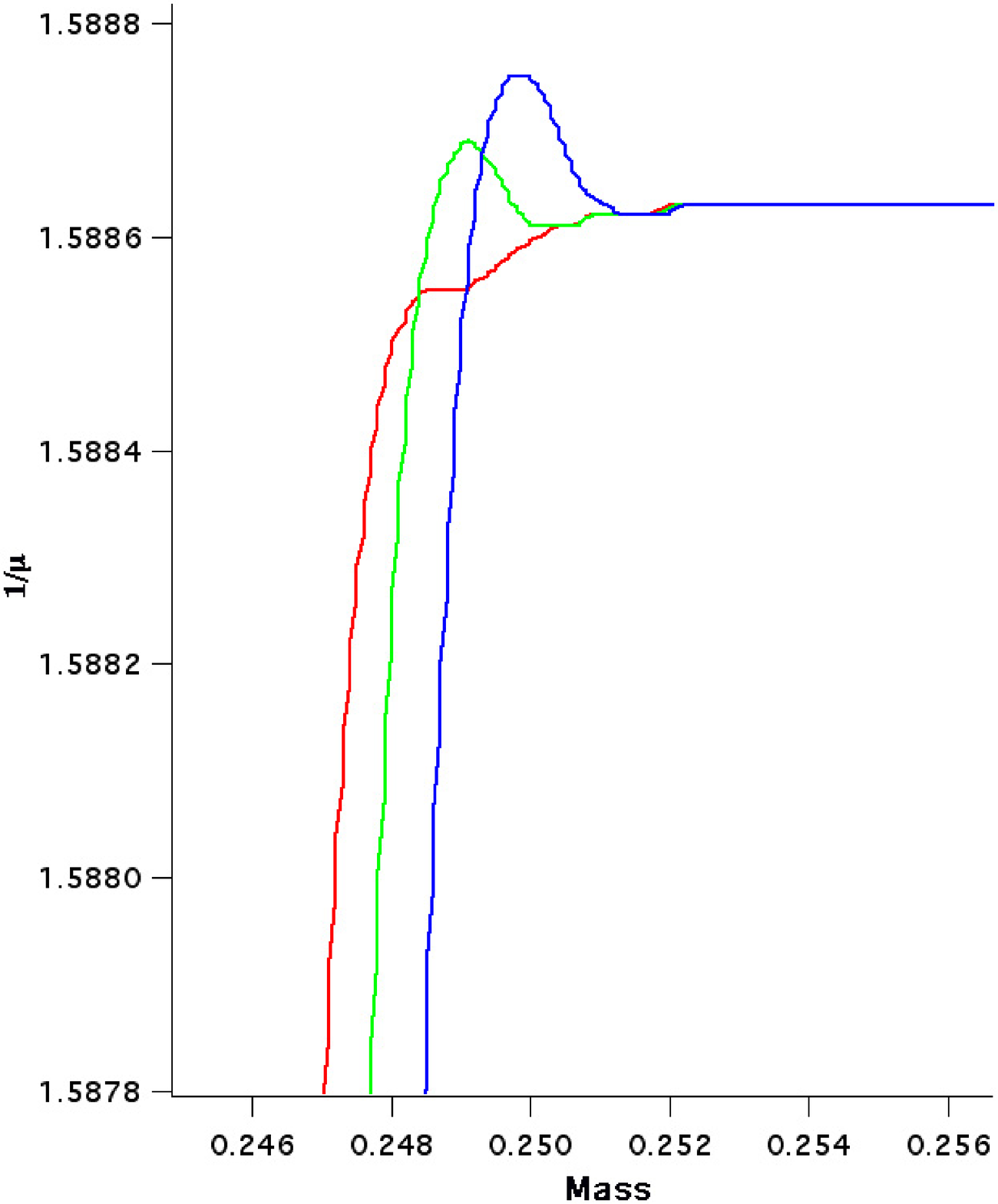,height=4.0in,bbllx=0pt,bblly=0pt,bburx=750pt,bbury=770pt,clip=}}
{\small {\bf Figure 3}. The profile of reciprocal molecular weight ($1/\mu$) 
as a function of mass in solar units, at three successive times (red, then 
green 2 Myr later, then blue 2Myr later still). } 
\vskip 0.1truein

\par At a stage (Fig. 3) when the SCZ has only just begun to 
retreat, there is no bump in $1/\mu$, but just a slight
distortion at about 0.248$\Msun$. This is because the \het\ consumption is
taking place in a region where there is still a substantial $\mu$ gradient
left over from earlier history. But as the H-burning shell moves out (in 
mass), the \het-burning shell preceding it moves into a region of more uniform 
\ho/\hef\  ratio, and so the peak in $1/\mu$ begins to stand out.
By the time the leading edge of the shell has moved to $0.25\Msun$ there is a 
clear local maximum 
in $1/\mu$, which persists indefinitely as the H-burning shell 
advances and the convective envelope retreats.

\par At a point somewhat beyond this in the evolution of our 1D star (Fig. 1)
we mapped the 1D model on to a
3D model and used the hydrodynamic code `Djehuty' developed at the Lawrence 
Livermore National Laboratory ({\it 13, 14, 15}). The code is described most 
fully in the third of these papers. Although Djehuty is designed to deal with 
an entire star, from center to photosphere, we economised on meshpoints by 
considering only the region below the SCZ. It is important for numerical 
purposes that the 1D and 3D codes use exactly the same approximations for 
physical processes, e.g. equation of state, nuclear reaction rates, opacities.
\par The location of the starting model of the 3D calculation is shown on
Fig. 1. If we had been clear before starting the 3D calculation that the
$1/\mu$ bump was going to cause mixing, we would have started further down,
at the point where the bump first presents itself, which is just above the 
point labelled `deepest penetration'. It has become clear that our unexpected
mixing will begin around here, and in practice we expect (as discussed
below) that almost all of the \het\ in the SCZ will have been consumed by the
time the model reaches the point where our 3D calculation started. Since 3D
modeling is very expensive of computer time, we have chosen not to redo the 
calculation for an earlier starting-point. Fig. 4 is a cross-section of the
starting model for the 3D run, and shows the $\mu$-inversion as a ring
well outside the burning shell.

\vskip 0.1truein
\centerline{\psfig{figure=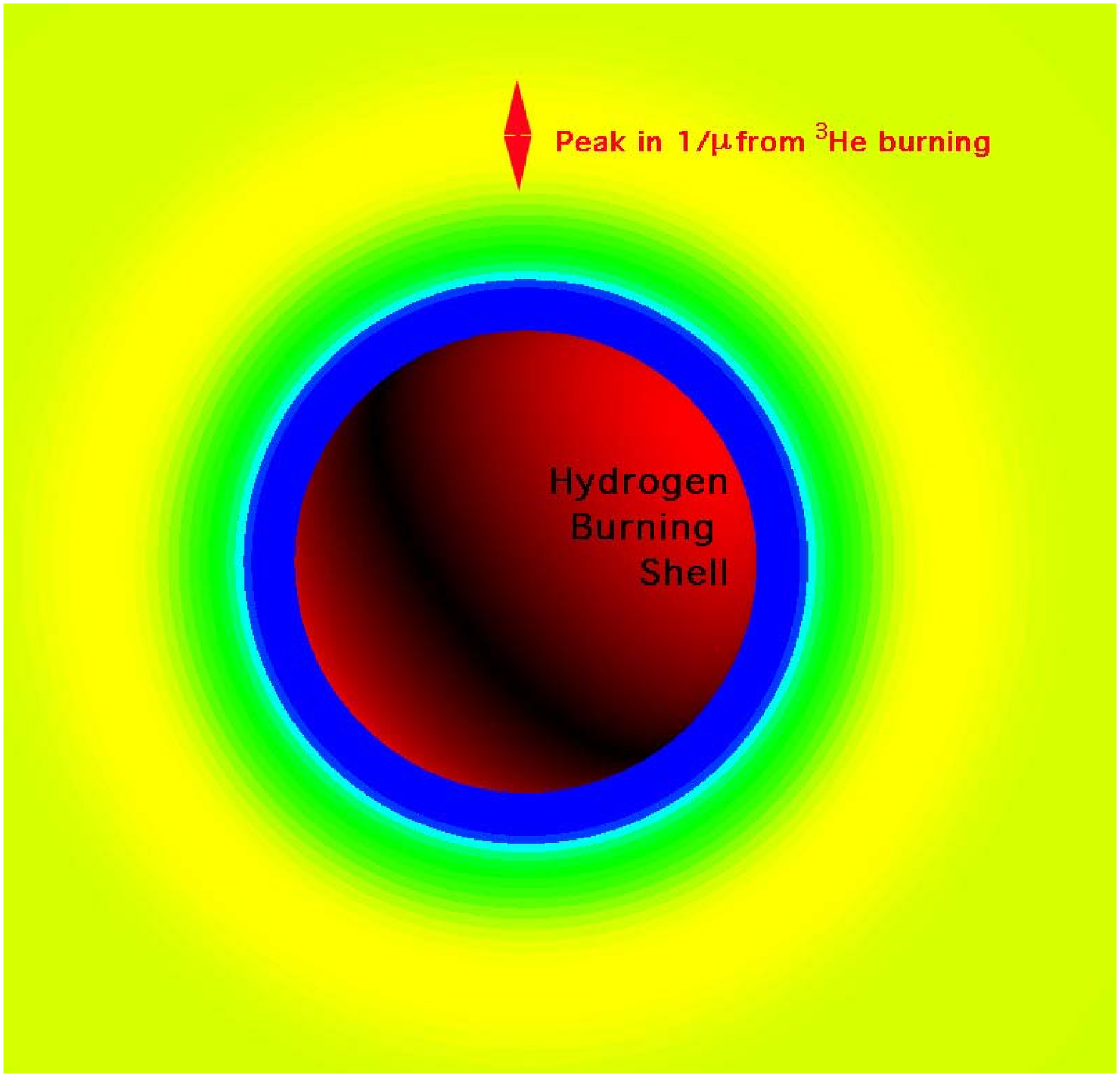,height=4.5in,bbllx=0pt,bblly=0pt,bburx=780pt,bbury=770pt,clip=}}
\vskip 0.1truein 
{\small {\bf Figure 4.} A color-coded plot of $\mu$ on a cross-section through 
the initial 3D model. The shell where the $\mu$-inversion occurs is the yellow 
region sandwiched between a yellow-green and a rather darker green. The 
inversion is at a radius of
$\ts 5.10^7\thn$m. The base of the SCZ is at $\ts 2.10^9\thn$m, well outside
the frame, and the surface of the star is at $\ts 2.10^{10}\thn$m.} .
\vskip 0.1truein 

\par Following the early development of the initially-spherical shell on
which $1/\mu$ has a constant value near its peak (Fig. 5), the surface has 
begun to dimple after only $\ts 800\thn$secs, and by $2118\thn$secs the 
dimpling is very
marked and the surface has begun to tear. Some points have moved $\ts 2\%$
radially, i.e. $\ts 10^6\thn$m, indicating velocities of $\ts 500\thn$m/s. 
The mean velocity decreases slightly in the passage from the second to the 
fourth panel. Other spherical shells, well away from the inversion on either 
side, show no such dimpling, at least until the influence of the inversion has 
spread to them. A movie of which Fig. 5 is four frames is given as Movie.S1 in
Supplementary Online Material. 

\vskip 0.1truein
\centerline{\psfig{figure=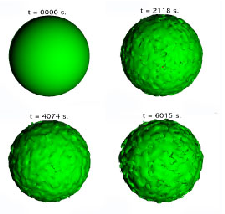,height=4.5in,bbllx=0pt,bblly=0pt,bburx=60pt,bbury=57pt,clip=}} 
\vskip 0.1truein
{\small {\bf Figure 5.} The development with time of a contour surface of mean 
molecular weight near the peak in the blue curve of Fig. 3. The contour 
dimples, and begins to break up, on a timescale of only $\ts 2000\thn$sec. }  
\vskip 0.1truein

\par The velocity we see is roughly consistent with the expectation
that it should be $v^2\ts g l \dm/\mu$, where $g$ is the local gravity and $l$
is the local pressure scale height.
The motion appears turbulent, and has the effect of diluting the
inverse molecular-weight gradient, but it cannot eliminate it. As the
turbulent region entrains more of the normally stable region outside it yet
below the normal convective envelope, it brings
in fresh \het, which  burns at the base of this mixing region, thus sustaining
the inverse molecular-weight gradient. Ultimately this turbulent region will
extend to unite with the normally-convective envelope, 
so that the considerable reservoir of \het\ there will also be depleted.
If its speed of $\ts 500\thn$m/s is maintained the time for processed material 
to reach the classically unstable SCZ is only about 1 month, while the time 
for the H-burning shell to 
burn through the  $\ts 0.02\Msun$ layer is over $10^6\thn$yrs.
\par The above argument establishes that the mixing in the SCZ is extended 
below the
classical convective limit, and that it is very fast compared to the nuclear
timescales of either the hydrogen-burning shell or the \het-burning reaction. 
We estimate from the nuclear-burning rates we find that as the hydrogen
shell burns outwards
the \het\ will be destroyed in 16 times as much mass as the hydrogen shell
burns through.

\par We believe that the extra mixing that we have seen gives a
satisfactory answer to the problem of matching the \het\ abundance of
Big Bang nucleosynthesis. Although low-mass stars do indeed
produce considerable amounts of \het\ on the MS, this will all be
destroyed by the substantially deeper mixing that we now expect on the RGB.
\par Our deeper mixing can also be relevant to further problems. 
According to the classical models
of RGB stars, there is no further modification to the composition in an RGB
convective envelope after it has reached its maximum extent early on the RGB. 
Yet observations persistently suggest that the ratios $^{13}$C/$^{12}$C and 
$^{14}$N/$^{12}$C both increase appreciably as one goes up the RGB 
({\it 16, 17}). Both
these ratios can be expected to increase only if the material in the envelope
is somehow being processed near the H-burning shell. Our model makes this very
likely. Although the $\mu$-inversion that we find is somewhat above the main
part of the H-burning shell, it is not far above and we can expect some modest
processing of $^{12}$C to $^{13}$C and $^{14}$N. According to ({\it 18}),
it appears to be necessary for some extra mixing to take
place beyond the point on the RGB where the SCZ has penetrated most deeply; 
that is exactly the point where our mechanism should start to operate. In 
({\it 18, 19, 20, 21}) it was suggested that rotation in the region between 
the SCZ and the hydrogen-burning shell might be responsible for the required 
mixing. We do not dispute the possible importance of rotation; however we 
emphasise that the mechanism we have discovered is not ad hoc, but simply 
arises naturally when the modeling is done in 3D. This mixing occurs regardless
of possible variables like rotation and magnetic fields. It seems possible 
to us that different rates of rotation might vary the efficiency of our 
process, and we intend to investigate models with rotation in the future.
\par Correlations between abundance excesses and deficits of various elements
and isotopes in the low-mass evolved stars of globular clusters have been 
reported in ({\it 16}). Although it is hard to distinguish star-to-star 
variations that may be due to evolution from those that may be due to 
primordial variation, we expect our mechanism to lead to substantial 
evolutionary variations.

\par We feel that our investigation demonstrates particularly
clearly the virtue of attempting to model in 3D, where the motion evolved
naturally, and to a magnitude that initially surprised us. 

{\bf References and Notes}
\begin{enumerate}
\def\z{\ \ \thn } 
\item  D. S. P. Dearborn, in {\it The Sun in Time}, C. Sonett, M. Giampapa, M. Mathews, Eds. 
ISBN 0-8165-12987-3 University of Arizona Press, Tucson, AZ, USA p159 (1991)
\item  P. P. Eggleton, {\it MNRAS}  {\bf 156}, 361 (1972).
\item  O. R. Pols, C. A. Tout,  Zh. Han, P. P. Eggleton, {\it MNRAS} {\bf 274}, 964 (1995)
\item  I. Iben, Jr, {\it Astrophys. J.} {\bf 147}, 624 (1967)
\item  D. S. Balser, T. M. Bania, R. T. Rood, T. L. Wilson {\it Astrophys. J.} 
510, 759 (1999). The initial value for \het\ that we assumed is somewhat higher than the
mass-fraction ($\ts 5.10^{-5}$) implied by this reference. This is partly because 
we assume that primordial deuterium,
of comparable abundance, is wholly burnt into \het\ before the computation 
starts. However the important point is that the great bulk of the \het\ in
the RG phase is what was synthesised from ordinary hydrogen during the MS
phase, and not what was there initially. The enrichment factor of 8 that we 
mention above would be a factor
of $\ts 16$ if we started with half as much \het, but the abundance level of
$\ts 1.6.10^{-3}$ would be very much the same.

\item  J. Faulkner, {\it Astrophys. J.} {\bf 144}, 978 (1966)
\item  J. Faulkner, {\it Astrophys. J.} {\bf 173}, 401 (1972)
\item  G. Steigman, D. S. P. Dearborn, D. Schramm, in
 {\it Nucleosynthesis and its implications on nuclear and particle physics}, J. Audouze,
 N. Mathieu, Eds {\it NATO ASI Series}. Volume C163, p37 (1986) 
\item  D. S. P. Dearborn, D. Schramm, G. Steigman, {\it Astrophys. J.} {\bf 302}, 35 (1986)  
\item  D. S. P. Dearborn, G. Steigman, M. Tosi, {\it Astrophys. J.} {\bf 465}, 887 (1996) 
\item  N. Hata {\it et al.},  {\it Phys. Rev. Lett.} {\bf 75}, 3977 (1995)
\item  K. A. Olive, R. T. Rood, D. N. Schramm, J. Truran, E. Vangioni-Flam, {\it Astrophys. J.} {\bf 444}, 680 (1995) 
\item  G. Baz{\'a}n {\it et al.}, in {\it `3-D Stellar Evolution'} S. Turcotte, 
 S. C. Keller, R. M. Cavallo, Eds ASP conf. {\bf 293}, p1 (2003) 
\item  P. P. Eggleton {\it et al.}, in {\it `3-D Stellar Evolution'} S. Turcotte,
S. C. Keller, R. M. Cavallo, Eds ASP conf. {\bf 293}, p15 (2003)
\item  D. S. P. Dearborn, J. C. Lattanzio, P. P. Eggleton, {\it Astrophys. J.} 
 {\bf 639}, 405 (2006)
\item  N. Suntzeff, in {\it The globular clusters-galaxy connection},  G. H. Smith, J. P. Brodie, Eds {\it ASPC} {\bf 48}, 167 (1993)
\item  R. P. Kraft, {\it PASP}  {\bf 106}, 553 (1994)
\item  A. Weiss, C. Charbonnel, {\it Mem. S. A. It.} {\bf 75}, 347 (2004)
\item  A. V. Sweigart, K. G. Mengel, {\it Astrophys. J.}  {\bf 229}, 624 (1979)
\item  C. Charbonnel,  {\it Astrophys. J.} {\bf 453}, L41 (1995)
\item  P. A. Denissenkov, C. A. Tout, {\it MNRAS} {\bf 316}, 395 (2000)
\item  We are indebted to R. Palasek for managing the code, and for assistance
with the graphics. This study has been carried out under the auspices of the 
U.S. Department of  Energy, National Nuclear Security Administration, by 
the University of  California, Lawrence Livermore National Laboratory, 
under contract  No. W-7405-Eng-48.
\end{enumerate}
\pn{\bf Supporting Online Material}
\pn www.sciencemag.org
\pn Movie S1

\vfill
\eject

\vskip 0.2truein
{\small {\bf Figure 1}. Evolution of a low-mass Pop I star in a 
luminosity-temperature diagram. The model was computed in 
1D, i.e. spherical symmetry was assumed, using the code of ({\it 1, 2})
with updated equation of state, opacity and nuclear reaction rates ({\it 3}). 
Surface temperature is in Kelvins, luminosity in Solar units.}

\vskip 0.1truein
\vskip 0.2truein
{\small {\bf Figure 2}.  (a) Profiles of the abundances of certain nuclei in 
a star which has evolved to roughly the end of the MS (Fig. 1; 
$T\ts 5000\thn$K, $L\ts 2\Lsun$). $^1$H is orange, $^4$He is red,
$^{16}$O is blue, $^{12}$C is black, $^{14}$N is green and \het\ is yellow.
\het\ shows a major peak where the abundance reaches $\ts 18$ times 
the initial (surface) abundance. (b) The same star later, when the SCZ 
reaches its maximum inward extent (Fig 1).
The \het\ peak has been homogenized, to a factor of 8 larger than its initial
value. The inert, H-depleted core is about $0.27\Msun$.}
\vskip 0.1truein

\vskip 0.2truein
{\small {\bf Figure 3}. The profile of reciprocal molecular weight ($1/\mu$) 
as a function of mass in solar units, at three successive times (red, then 
green 2 Myr later, then blue 2Myr later still). } 
\vskip 0.1truein

\vskip 0.1truein
\vskip 0.1truein 
{\small {\bf Figure 4.} A color-coded plot of $\mu$ on a cross-section through 
the initial 3D model. The shell where the $\mu$-inversion occurs is the yellow 
region sandwiched between a yellow-green and a rather darker green. The 
inversion is at a radius of
$\ts 5.10^7\thn$m. The base of the SCZ is at $\ts 2.10^9\thn$m, well outside
the frame, and the surface of the star is at $\ts 2.10^{10}\thn$m.} .
\vskip 0.1truein 

\vskip 0.1truein
\vskip 0.1truein
{\small {\bf Figure 5.} The development with time of a contour surface of mean 
molecular weight near the peak in the blue curve of Fig. 3. The contour 
dimples, and begins to break up, on a timescale of only $\ts 2000\thn$sec. }  
\vskip 0.1truein
\end {document}